\documentclass[sigconf,screen]{acmart}

\usepackage{subcaption}
\AtBeginDocument{%
  \providecommand\BibTeX{{%
    \normalfont B\kern-0.5em{\scshape i\kern-0.25em b}\kern-0.8em\TeX}}}

\copyrightyear{2022}
\acmYear{2022}
\setcopyright{acmlicensed}\acmConference[PDC 2022 Vol. 1]{Participatory Design Conference 2022: Volume 1}{August 19-September 1, 2022}{Newcastle upon Tyne, United Kingdom}
\acmBooktitle{Participatory Design Conference 2022: Volume 1 (PDC 2022 Vol. 1), August 19-September 1, 2022, Newcastle upon Tyne, United Kingdom}
\acmPrice{15.00}
\acmDOI{10.1145/3536169.3537781}
\acmISBN{978-1-4503-9388-1/22/08}

\begin{document}

\title{Empowering Participation Within Structures of Dependency}

\author{Aakash Gautam}
\affiliation{%
  \institution{San Francisco State University}
  \city{San Francisco}
  \country{United States}
  }
\email{aakash@sfsu.edu}

\author{Deborah Tatar}
\affiliation{%
  \institution{Virginia Tech}
  \city{Blacksburg}
  \country{United States}
  }
\email{dtatar@cs.vt.edu}


\begin{abstract}
% Participatory Design (PD) has long been concerned with political change to support people to have democratic control over the processes, solutions, and, in general, matters of concern to them. 
% A particular challenge remains in supporting vulnerable groups gaining power and control when they are dependent on organizations and external structures. 
% Challenging the organizations or structures may not be pragmatic given the dependency of the vulnerable group and limitations of our resources and capacities. 
% In this paper, we reflect on our five year engagement with survivors of sex trafficking in Nepal and an anti-trafficking organization that supports the survivors. 
% Arguing that the prevalence of deficit perspective in the setting promotes dependency and robs the survivors' agency, we sought to bring change by exploring possibilities based on the survivors' existing assets. 
% We present three configurations that illuminate how our design decisions and collective exploration operate to empower participation while attending to the substantial power implicitly and explicitly manifest in existing structures. 
% We highlight the challenges we face and contend the need for long-term, incremental engagement to promote democratic control.

%%%%%%%%%%%%%%%%%%%%%%%%%%%%%%%%%%%%%%%%%%%%%%
% Nothing about agonism in the abstract.

%%%%%%%%%%%%%%%%%%%%%%%%%%%%%%%%%%%%%%%%%%%%%%

Participatory Design (PD) seeks political change to support people's  democratic control over processes, solutions, and, in general, matters of concern to them. A particular challenge remains in supporting vulnerable groups to gain power and control when they are dependent on organizations and external structures. We reflect on our five-year engagement with survivors of sex trafficking in Nepal and an anti-trafficking organization that supports the survivors. Arguing that the prevalence of deficit perspective in the setting promotes dependency and robs the survivors' agency, we sought to bring change by exploring possibilities based on the survivors' existing assets. Three configurations illuminate how our design decisions and collective exploration operate to empower participation while attending to the substantial power implicitly and explicitly manifest in existing structures. 
We highlight the challenges we faced, uncovering actions that PD practitioners can take, 
including an emphasis on collaborative entanglements, attending to contingent factors, and encouraging provisional collectives.

%We highlight the challenges we face and contend the need for long-term, incremental engagement to promote democratic control.
\end{abstract}



\maketitle

\section{Introduction}

A central tenet of all PD engagements is to support people's voices in the design endeavor, offering opportunity for them to enact their desired futures \cite{bratteteig2016participatory, nygaard1975trade, simonsen2012routledge, bodker2018participatory}. 
The emphasis on political participation, highlights the importance of agency-building and empowerment of participants in PD (e.g., \cite{beck2002p, gautam2020p}). 
To this end, calls have been made to (re)politicize design, confront the  ``big issues'' \cite{huybrechts2020visions, bodker2018participatory} and adapt agonistic approaches \cite{bjorgvinsson2012agonistic, disalvo2015adversarial}.
At the same time, scholars have argued for attending to relations and interactions, configuring the micro-elements of design to the local context where design is situated \cite{light2014structuring, akama2018practices, light2012human, clarke2019socio, akama2010community}. 
These two seemingly different positions are closely related: both necessitate that PD practitioners be reflexive about our role and action, pay attention to the visions we aspire to achieve, and engage in actions that enable egalitarian participation and power-sharing in the context \cite{simonsen2012routledge, light2018nature, bjorgvinsson2010participatory, ehn2017learning, bratteteig2016participatory, bodker2018participatory, huybrechts2020visions}.

In this paper, we contend that PD's role in politicization which involves the ``articulation of divergent, conflicting, and alternative trajectories of future ... possibilities and assemblages'' \cite{swyngedouw2007post} requires care, particularly in contexts where conflict may place already-vulnerable groups at a position of greater harm. 
We reflect on our work with a Nepali anti-trafficking organization and the sex-trafficking survivors who live in the organization's shelter homes.
The survivors depend on the organization for support during their journey of achieving, what they call, ``dignified reintegration'' into Nepali society.  
Financial independence, social acceptance, and agency in dealing with societal actors and institutions are all components of dignified reintegration.
We aim to support the survivors in positions of greater power in their reintegration journey. 
% This involved supporting the survivors in developing and using their knowledge and skills we believe can endure beyond the shelter home. 
% Our work also involves exploring ways in which the survivors could be at a position of greater power within the shelter home as it is a significant stage in their reintegration journey. 

Survivors are in a fragile position. While not all survivors of sex-trafficking rely on anti-trafficking organizations, many do. 
Survivors of trafficking are typically from poor families and are illiterate \cite{nhrc2018}. Post repatriation, many are shunned by their families. 
Most survivors in the organizations' care are women\footnote{Men and transgender people are also trafficked but due to deeply rooted patriarchy, the dominant discourse in Nepali society sees only women as survivors of trafficking. A few organizations, including our partner organization, have started working with men and transgender people. However, the available services are extremely limited.}
who, due to deeply ingrained patriarchal beliefs, face societal stigma as well as economic and political challenges \cite{kaufman2011research, crawford2008sex, joshi2001cheli, poudel2011dealing, mahendra2001community, laurie2015post, simkhada2008life, sharma2015sex, richardson2016women}.
%Many survivors rely on anti-trafficking organizations as a source of support throughout their reintegration journey.

Aside from these realities, factors within the anti-trafficking organization may contribute to survivors' precarious position. Our research identified an overarching approach in which survivors are positioned as passive recipients of services provided by the organizations, tied to deep beliefs about the survivors' deficiencies.
There are two major factors behind this portrayal: very limited regulations with no guidelines or best practices, and lack of resources \cite{gautam2018social, gautam2020crafting}. 
Neither of the organizations where we conducted our initial study had a clear measure of what constituted a successful reintegration \cite{gautam2018participatory}. 
Without examples of best practices, the organizational processes embodied the deficit-based beliefs held by staff members (and, to some extent, the survivors). Further, anti-trafficking organizations in Nepal receive limited support from the government, forcing them to compete with one another for grants, a significant proportion of which are funded by a small group of international organizations. These international organizations respond to grant proposals that emphasize survivor needs and deficiencies. The limited number of opportunities invites conservative attitudes towards conceptualizing and writing grant proposals. 
% To gain funding, anti-trafficking organizations have to design and propose projects that align with the donors' priorities which may not align with the survivors' needs. 

We begin with the belief that survivors' power and agency within the shelter home can be transferred outside of it, facilitating their dignified reintegration.
The survivors' position as passive recipients of support denies them agency and a voice within the shelter home. 
As a result, we believe that one aspect of our involvement should be to facilitate more democratic control over the organization's processes, with survivors' voices at the center.
The change is political and requires us to side with the survivors, leveraging our power and position of privilege as researchers.

Yet, the reality of dependence leads us to take incremental steps with buy-in from the organizational partner. 
Direct confrontation on changing the organization's processes is not viable for several reasons.
First, we have an obligation to the organization to work together and need to continue to earn their trust and belief.
Our access to the survivors depends on the organization trusting us.
Negotiating change in established processes without forming a prior relationship and proven evidence that the alternative works is challenging.  
Second, and perhaps more critically, the existing system, despite its problems, provides real support to the survivors; it would be wrong to discount the value of the organization's resources, especially since we have little or nothing to offer if our proposed approaches fail.
%Our political move is guided by our ethical concern that we do not want to cause harm. 

%The prevalence of vulnerable group's deficits held by the organizations as well as the vulnerable group themselves adds to the challenge. 

We shaped our approach away from focusing on the survivors' deficits, focusing on the knowledge, skills, and resources---their assets---available to the survivors\footnote{We call the survivors we worked with ``sister-survivors'' and use ``survivors'' to denote the larger group of survivors of trafficking.}. We define assets as those strengths, attributes, and resources that can be brought into relevance to satisfice the inherent tensions between a member of a population's needs, their understood or experienced aspirations, and the structural limitations of the system.
Success, thus, is constituted by the artful integration \cite{suchman2002located} of these concerns into action that (1) considers the structural limitations to ensure that, at minimum, does no harm, (2) leads to positive experiences for the sister-survivors and the organization, (3) offers potentially useful learning that finds some uptake, and (4) puts the sister-survivors in a position to use their assets and exercise agency.

The focus on the assets, or assets-based design, intends to show examples of an alternative approach, enabling us to get buy-in from our organizational partner. 
% We discuss three configurations to empower within the structure that attend to different levels of structural limitations and dependency present in the setting while exploring moves to empower the sister-survivors. 
We discuss three configurations that can empower within the structure. Each attends to different level of structural limitations and dependency present in the setting. 
The three configurations involve (1) conceptualizing the relationship between the anti-trafficking organization and the sister-survivors as mutually dependent, (2) attending to the organization's technological aspirations, and (3) navigating the organization's dependence on donor funds to create space for the survivors' voices in the design of projects.
These configurations highlight details of the local context to describe how we arrived on the design approaches and how the decisions were made by ensuring that both the organization and the sister-survivors were involved, albeit not equally.
 
By sharing these configurations, we seek to answer, ``How can PD engagements push back on prevailing deficit perspectives and support vulnerable groups to gain power and control within the existing structure of dependency?''
We make a three-fold contribution. 
First, we present an assets-based design lens on PD. 
We contend that participatory approaches can be configured to emphasize assets and, in the process, support greater power and agency in vulnerable groups' participation with us. 
Second, we argue that assets-based approaches can illuminate alternative possibilities and futures. Assets-based design is a way to include vulnerable groups' voices, emphasizing their agency even in contexts where they are dependent on the organization.
%incrementally pushing for changes in the structure they are dependent upon.
Third, we make a case for taking incremental steps, bridging the move to agonistic design and change by showcasing examples of alternative approaches---here, a focus on sister-survivors' assets---to allow space and time for organizations working with vulnerable groups to evaluate and embrace design propositions. 
We highlight the value of promoting collaborative entanglement, attending to contingencies, and emphasizing a provisional collective in balancing the multiple tensions that are present in contexts with structural constraints and dependencies.

\subsection{Reflexivity and Positionality}
% PD scholars should “reflect critically on their skills, their motivations, their practices, their relation- ships, and their priorities” \cite[pp. 9]{dearden2008participatory}. 
% ``We need to bear in mind that there are no innocent positions (Haraway 1988), not even for researchers.'' \cite[pp. 87]{karasti2010taking}
% Further, Karasti urges us to move away from focusing` `on the ‘other’ (technology and user) at the expense of concern for the self and relationships.'' \cite[pp. 89]{karasti2010taking}.
% In addition to a researcher’s personal self-examination, interpersonal and institutional reflexivity are also required in PD (e.g., Blomberg et al. 1996; Suchman et al. 1999; Balka 2006).

Dearden and Rizvi argue that PD scholars ``must reflect critically on their skills, their motivations, their practices, their relationships and their priorities'' \cite[pp. 8]{dearden2008participatory} (see also \cite{brule2019negotiating, liang2021embracing}). 
Indeed, our motivation, skills, relationships with organizations in Nepal, and our position in academic setting in the US have influenced the project. 

This project began with an investigation into the potential for design to alter power and authority.
The first author is a Nepali male from a relatively privileged background. 
Growing up, he witnessed the impact of gender inequality in Nepali society and, through his research, sought to investigate ways to reduce gender inequality, which served as the motivation for the work.
The second author, a female US citizen and the first author's advisor, has been involved throughout the project. 
Her work has addressed issues of design and power in the workplace and classroom.
We both find the pervasive patriarchal values prevalent in Nepali society problematic, and see PD playing a role in challenging deeply ingrained beliefs. 

We are associated with the computer science department in our institutions. 
%Our skills involve designing socio-technical systems, so our focus has been on attending to the social and technical configurations in the setting. 
Our technical expertise facilitated our acceptance as collaborators with Nepal's two anti-trafficking organizations.
However, sustainable change cannot be realized with only technical solutions; potential solutions have to attend to social issues. 
We focus on both the social and the technical. We see the sensitive construction of technical configurations as important for the organization, the prospects of the sister-survivors, and for the reconfiguration of human-computer interaction to center the social.
We do not hold an ``innocent position'' \cite{haraway1988situated}; instead, we seek to leverage our position and power, including our technical skills, to center the survivors' voices, and more broadly, deal with the political issues to realize greater equity. 

We envision the survivors in positions of power and authority in the organization and in society as a whole. 
While we wish for radical social change, we are aware of our personal and institutional limitations.
The context is delicate, necessitating extensive justification for our actions.
We have limited resources to support the survivors if our move fails. 

Our approach has to be sustainable and ensure that it, at the bare minimum, causes no harm to an already-vulnerable group. 
In this regard, we have sought to collaborate with existing institutions on the ground---the anti-trafficking organizations---to find ways to move forward together. While we are aware of some of the issues surrounding anti-trafficking efforts in Nepal \cite{kaufman2011research, laurie2015post, crawford2008sex}, in the lack of governmental and other institutional support, we see the services provided by anti-trafficking organizations as central to many survivors of trafficking.
Developing a trusting and mutually beneficial relationship with the anti-trafficking organization has been critical to our work.
To that end, we have been supporting the group by performing tasks such as maintaining their website and raising funds for a shelter home during the initial wave of the COVID-19 outbreak.

\section{Power Within Structures of Dependency}

Power lies at the heart of PD. 
In their goal to promote democratic control, PD practitioners examine where the locus of power lies and strive to collectively design interventions to promote sharing of power among the users and other stakeholders \cite{simonsen2012routledge, ehn2008participation, bratteteig2016participatory, muller2009participatory, clement1993retrospective, charlotte2020decolonizing}.
Historically, Nygaard's engagement with the Norwegian Iron and Metal Workers Union (NJMF) highlighted a set of methods and practices for supporting workers' power through their local knowledge in negotiating and making decisions on issues concerning ``new technologies'' \cite{ehn2017learning, simonsen2012routledge, ehn2008participation}.
While PD expanded beyond supporting changes at the workplace to home and broader community-based transformations, the examination of power enacted in its various forms remains central \cite{bratteteig2012disentangling,simonsen2012routledge,le2016design,dickinson2019cavalry,harrington2019deconstructing,bjorgvinsson2010participatory,kendall2020politics}.

Power is multi-faceted and dynamic \cite{mills2003routledge, allen2005feminist}.  It ``circulates and operates in the form of a network permeating through the various levels of the system'', and is ``enacted and actively contested among various agents in a  system'' \cite[pp. 2]{kannabiran2010politics}. 
PD scholarship highlights several challenges in examining the locus of power and in bringing change in the power dynamics \cite{light2012human, bjorgvinsson2010participatory, simonsen2012routledge, bratteteig2012disentangling, gautam2018participatory, akama2010community}.
Bratteteig and Wagner \cite{bratteteig2012disentangling}, building on Borum and Enderud \cite{borum1981konflikter}, call attention to four mechanisms whose various arrangements make the enactment of power less visible and, thus, difficult to challenge: (1) agenda control, (2) participants, (3) scope, and (4) resources. 
PD endeavors seeking to empower the vulnerable and the marginalized need to attend local enactments of these mechanisms that seek to resist change in power dynamics. 
As such, we echo PD scholars (e.g., \cite{akama2018practices, andersen2015participation, vines2013configuring, charlotte2020decolonizing, kendall2018disentangling}) in arguing \emph{against} a universal standard for participation and instead push for PD to be configured according to the local context.
Attending to how participation is configured remains central across various domains where profound inequities are present such as in health \cite{braa2012health, byrne2007participatory, davidson2013health},  education \cite{disalvo2016participatory, wong2020culture, dindler2020computational}, civic participation \cite{dickinson2019cavalry, asad2017creating, devisch2018participatory}, and  broader community engagement \cite{costanza2020design, mainsah2012social, le2015strangers, thinyane2018critical, kendall2018disentangling}.
In this regard, PD practitioners play an influential role (e.g., \cite{light2012human, harrington2019deconstructing, bodker2018participatory, clarke2019socio, bjorgvinsson2012agonistic, disalvo2015adversarial, dantec2013infrastructuring, karasti2004infrastructuring, charlotte2020decolonizing}).
After all, how we set up the participatory engagement, what materials and resources we use in the process, how we define participation, and how we position different stakeholders affects what issues are centered and how local power dynamics receives attention.

% Akama and Light present ``PD as configured by the people, practices, place, and structures with which it is entangled, rather than arguing for universal gold standards for participation'' \cite[pp. 1]{akama2018practices}. We endorse this belief. We critically ask, ``What does it mean to configure PD in a context where dependency is present and cannot be challenged?''

%Agonism also admits that power and exclusion can never be completely erased, so they must be made visible and contestable. Such egalitarian vision and embrace of contestation in agonism require permanent spaces for conflicts and confrontations, where various members of the public engage in passionate disagreements and dialogues that tinker with orders and hegemonies not rooted in their conceptions of the good.

\subsection{Bridging Agonism with Assets}
\label{sec:bridge}
% What decisions are made, who makes it, and why---all of these involve the enactment of power \cite{bratteteig2012disentangling, simonsen2012routledge}.

Recent work has critiqued PD engagements for being limited to ``do-gooding'' where PD is seen as a good in and of itself \cite{bodker2018participatory, huybrechts2020visions}.
Instead, scholars have made a call to move away from setting an agenda that focuses only on the here-and-now towards embracing the politics and conflicts that are inherent when seeking to promote democratic control and change  \cite{bodker2018participatory, huybrechts2020visions, beck2002p, karasti2010taking}. 

One such approach to contesting power dynamics involves PD practitioners designing agonistic approaches where ``the hegemony of dominant authority is potentially challenged through manifold forceful but tolerant disputes among passionately engaged publics'' \cite[pp. 128]{bjorgvinsson2012agonistic}. 
Agonism is comprised of a commitment to enabling multiple perspectives and values, establishing spaces where forever-and-ongoing contestation can occur between members of the publics, and holding a belief that such conflict is necessary and helpful in fostering an ethos of democratic engagement, particularly to challenge what has been accepted as the ``natural'' order \cite{mouffe2009democracy, wenman2013agonistic} (see also \cite{hillgren2016counter, disalvo2015adversarial, bjorgvinsson2012agonistic, korn2015creating}). 
DiSalvo posits three tactics for agonism: revealing the hegemony of forces influencing the setting, reconfiguring aspects of what has been excluded and seeking to include it, and articulating a collective that challenges or provides alternatives to dominant practices \cite{disalvo2015adversarial}.   
We endorse the need for agonism in making existing power dynamics visible \emph{and} contestable, and, with it, enabling democratic control.

In our context, the prevalence of beliefs about the survivors' deficits and needs eroded both the survivors' self-belief and also the organization's beliefs about what the survivors can do.  
The institutionalization of this kind of deficit perspective has led scholars to characterize anti-trafficking organizations, including those in Nepal, to be part of a ``rescue industry'' \cite{agustin2007sex, laurie2015post}.
Agonistic approaches are needed to push back.

However, we remain cautious about embracing ``profound conflicts'' \cite{bodker2018participatory}. 
We see two major issues. 
The first issue revolves around power dynamics.
Agonism relies on the notion of adversaries which ``... characterize a relationship that includes disagreement and strife but that lacks a violent desire to abolish the other'' \cite[pp. 6]{disalvo2015adversarial}. 
Adversaries are conceptualized as on equal grounds in terms of power and agency, a condition that is challenging in contexts where one group is dependent on the other.
The second issue involves the potential repercussions of conflict, even when disputes are ``tolerant''.
There are serious costs in undertaking agonistic approaches \cite{bjorgvinsson2012agonistic}. 
It is higher for vulnerable groups who may not have support mechanisms if conflicts escalate.
Considering this, we argue that PD practitioners who are taking sides with the vulnerable group need to bridge the gap by making incremental moves, allowing for change to occur without threatening the existing structures of dependency. 
%In this paper, we present our configurations that carefully balanced the need to present alternative perspectives to push back on dominant belief while attending to the fact that confrontation may be detrimental to the well-being of an already-vulnerable group. 
Our particular configurations leverage the designers' position of power to amplify the voices of the vulnerable group and open consideration for alternative possibilities.

\subsubsection{Balancing Tensions on Taking a Side}

There is a wide consensus that participation is neither dichotomous nor homogeneous but rather involves varying degrees of engagement throughout the design process \cite{vines2013configuring, andersen2015participation, carroll2007participatory, winschiers2012community}.
Andersen et al. \cite{andersen2015participation} see participation as expressing ``matters of concern'' where ``agency is always derived from many interfering sources, rather than possessed by individuals'' and that  ``participation come into existence in various ways and situations'' \cite[pp. 251]{andersen2015participation}.
Central to this is the belief that participation has to be carefully formulated in PD engagements \cite{vines2013configuring, andersen2015participation, light2012human}. 
Not all stakeholders can be similarly involved, especially if we seek to center the voices of the vulnerable and marginalized in the design process (e.g., \cite{brereton2008new, kyng2015creating}).

PD practitioners can leverage their position to amplify the voices of the vulnerable \cite{akama2018practices, kendall2018disentangling}.
We used our position and power to amplify the sister-survivors' voices, first, in our design process and, subsequently, in the organization's processes. 
While we seek to take sides with the sister-survivors, we are aware that their well-being is dependent on the organization's ongoing support. 
It is also critical to align our approach with the local actors to ensure that the change is sustainable \cite{kendall2020politics, sultana2018design, kendall2018disentangling}.
Thus, we needed to balance two different goals \cite{tatar2007design}: one of reconfiguring existing practices so that the survivors are empowered, and another of ensuring that the moves are aligned with the organization's goals and priorities.
%In this regard, we promoted different levels of participation, moderating each to ensure that the survivors' voices are centered while also considering the staff members' perspectives in the design decisions.
%We also sought to illuminate examples from our engagement with the sister-survivors which we then presented to the organization as an alternative to their existing practices, and thus created space and time for the organization to evaluate our design proposition. 

\subsubsection{Focusing on Available Assets}

An assets-based approach ``centers the design process on identifying individuals’ and communities’ strengths and capacities and exploring feasible ways for users to build on these assets to attain desirable change'' \cite[pp. 2]{wong-villacres2021Reflections}.
PD can embody an assets-based orientation. 
In Brown et al. \cite{brown2018refugee} and Huybrechts et al.'s Traces of Coal project \cite{huybrechts2018building}, participants used their skills, strengths, and abilities to foster dialogue and act on bringing change. 
Moreover, mutual learning, a cornerstone of PD \cite{greenbaum1991design, simonsen2012routledge, muller2009participatory}, involves designers' reliance on participant expertise, i.e., their assets. 

In our case, we use PD methods to identify and build upon the sister-survivors' assets to realize two ends. 
First, to promote a ``shift in consciousness'' \cite{cornwall2016women, freire1970pedagogy}, a key tenet of empowerment. 
The shift enables people to engage with and challenge the ideas and practices that keep them in a state of perpetual subordination, and in the process, develop the capacity to bring about individual and institutional changes. 
The emphasis on assets aimed to encourage critical reflection among the sister-survivors, allowing them to see themselves as individuals and collectives with resources and strengths rather than passive recipients of assistance.
Second, we sought to bring everyone--the staff members and the sister-survivors--together as a collective to see value in the sister-survivors' existing assets and, with it, push back on the prevalent beliefs about the survivors' deficits.

\section{Methodology}

%%%%%%%%%%% New addition to make %%%%%%%%%%% 

% Focus on the messy bits of PD as Balka \cite{balka2010broadening} argues.
% Our position and the relationship we form in the setting matters \cite{karasti2010, balka2010} and highlighting these messy bits are valuable to PD scholarship. 
% These relationships influence the shape the project takes, including the kinds of configurations are seen as meaningful and valuable during the engagement. 
% Thus, it is important to reflect on how we believe we earned their trust, how we fostered a space for them to trust each other and us, and what enabled us to engage together towards agonistic design approaches. 

% Add part about how IRB was obtained and how consent was obtained with the survivors
% Describe what we did to make the sister-survivors comfortable in their participation with us. 

%%%%%%%%%%%%%%%%%%%%%%%%%%%%%%%%%%%%%%%%%%%% 

After several months of email correspondence and calls to establish a connection, we began our project in collaboration with two of the largest anti-trafficking organizations in Nepal.
Both organizations were interested in exploring technology to support their efforts. One of the organizations, who we refer to as ``Professional Organization'' (PO), had recently acquired a mobile application that enabled them to scan faces at the border and match them with faces on the missing person reports. 
Similarly, the other organization, who we call ``Survivor Organization'' (SO) since it was founded by survivors of trafficking and has many survivors on its staff, had recently concluded a two-week-long Photoshop and computer training program with a group of survivors.
Our technical expertise was influential in gaining access to both organizations.

The project involved three field studies and a remote study. 
The first field study (12/2017--01/2018) aimed to gain a comprehensive understanding of the two anti-trafficking organization's operations and the conditions surround the sister-survivors' lives in the shelter homes. This led us to identify critical assets available to the sister-survivors. 
In the second study (12/2018--01/2019), we examined the possibility of building upon those assets through a workshop on using computers. 
The third study (08/2019--10/2019) replicated and extended the second study by exploring possibilities for the sister-survivors to engage with societal actors and institutions by leveraging their assets.
The remote study (07/2020--10/2020) built on the sister-survivors' assets to make a case for including them in the organization's decision-making processes. 

Three factors resulted in the time gap between studies. 
First, we aspired for changes in belief within the organization and thus sought to engage the staff members in the design process. 
We analyzed data from one study, shared findings and plans with the staff members, and included their suggestions in the next iteration.  
Second, technology takes some time to implement. 
Third, the project was self-funded and the field studies took place at a time when the first author was able to leave his academic institution. 
Particularly, the first two studies were conducted over extended breaks between semesters.

By the end of the first study, we realized that we could not build a relationship of trust and a vision of shared goals with PO.
Trust is essential in ``designing value in the collaboration'' \cite[pp. 1]{warwick2017designing}.
Trust is also fragile. 
Distrust can form through interactions that are beyond the scope of our engagement. 
In our case, PO had earlier collaborated with researchers from the US who they felt had not been truthful about their objectives. 
A program officer at PO stated, ``\textit{... looks like we have had a history where even a professor from a reputed university came for a research and just used the organization for his research purpose and did not help it} [the organization] \textit{as promised.}''
We sought to repair trust by being transparent about our objectives, distancing ourselves from those US-based researchers, highlighting our background and positionality, and establishing trust based on the first author's volunteer experiences in PO's prior events. 
The approach was insufficient. 
PO expressed reservations, citing the fact that they have limited control over what is said about them \cite{alcoff1991problem}.
Thus, all of our subsequent engagements have been with SO only.

%% Commenting out the table of activities. 
%\input{activities_elaborate}

% Trust is central in a design engagement and researchers have to build trust before design research can begin \cite{clarke2019socio, yee2016goldilocks}. Indeed, in our context, we can see the need for trust as the partner organizations had to trust our and our approach despite the uncertainty of the outcome. 

\subsection{Partner Organization}
SO was formed in the late 1990's by a group of 15 women who were rescued and repatriated. 
Survivors of trafficking lead the organization. Many staff members across levels of the organization are survivors of trafficking. 
As of 2020, SO has more than 100 staff members employed across 14 districts in Nepal.  

%60.7\% of SO's annual funds in 2016 were from donor organizations.
Most anti-trafficking organizations in Nepal typically offer protected-living homes (shelter homes), skills-based training in handicrafts, and reintegration through the provision of jobs and/or reunification with families.
The skills-building program, in SO's case and in most anti-trafficking organizations in Nepal, involves training in creating local handicrafts such as glass-bead necklaces (\textit{Pote}), slippers, shoes, shawls, scarfs, and stone jewelry. 
Artifacts generated from the handicraft training are sold by SO as an additional source of revenue. 
%Both the organizations were reliant on donor organizations, most of which were international donor agencies. 
%Majority of the donor organizations prioritized programs that were highlighted by the US Department of State's annual Trafficking in Person's (TIP) report which provides an account of trafficking issues prevalent in the majority of the countries across the globe.

SO also received occasional support from local organizations. 
This enabled them to enroll five sister-survivors in a training program for trekking guides. In the third year of our engagement, all the sister-survivors were enrolled in a nearby ``morning'' school which offered classes from 6:30 to 9:30 am. These services are periodic, and the scope and duration depends on the sponsoring organization.

\subsection{Participants}

SO allows survivors to stay in the shelter homes ``as long as they need'', a policy which is more responsive to individual needs than the fixed 6-12 month duration system prevalent in most other NGOs. 
Survivors stay in SO's shelter homes anywhere from two months to six years. 
SO's and our programs have to be cognizant of this flux.

Over the past five years, we worked with around 35 sister-survivors between the ages of 13 and 23.
One had recently moved out of the shelter home and was renting an apartment in Kathmandu.
Only two had bank accounts. 
%SO managed the sister-survivors' finances, including providing for their basic daily needs and gave them 25\% of the proceeds from selling their handicrafts.
Around half could read Nepali text. Four reported being comfortable reading basic English text. 
None of the sister-survivors living in the shelter home had access to mobile phones or computers; most reported that they had never used a computer before.

\subsection{Ethics and Participant Comfort}

All of our studies were approved by our institution's Institutional Review Board (IRB). In all the protocols, verbal consents were obtained in Nepali and were phrased colloquially; nevertheless, we had to use words such as ``project'' and ``research'' which created a distance between us and the sister-survivors. IRB consent relies on individual agency which, in our context, could be difficult to enact especially considering that the sister-survivors often made collective decisions with the group. 
Throughout our engagements we reminded the sister-survivors that they were free to leave the study anytime and as a group they could opt out of the study. 

The IRB process, as currently structured, is more concerned with protecting the institution than with setting standards for ethical research.
We sought additional ways to promote comfort and control. 
For instance, some participants were always new to us. 
To promote comfort, the first author spent the week before each study visiting SO's office and the shelter home. He worked with the staff, but also introduced or re-introduced himself and the project to the sister-survivors, and engaged in discussions about other aspects of their lives, in a culturally appropriate way.
Only then did he invited the group to participate in further activities. 
The activities were also designed respecting the sister-survivors' desire to focus on the present and future rather than the past, and prioritized collective exploration over individual action.
\section{Three Configurations to Empower Within the Structure}

Our work seeks to challenge the prevalent deficit views and push for changes in organizational practices such that the survivors have a greater voice in the services that are designed for them. 
We leveraged our position in configuring our approach to balance multiple tensions present in the setting and make moves that helped us realize the changes (see Table \ref{tab:configurations}). 
%In particular, these configurations allowed us to explore an alternative approach of identifying and building upon the sister-survivors' existing assets. 
%They aim to illuminate a sustainable pathway to appreciate the sister-survivor's assets, and, thereby, bring about change in individual beliefs and organizational processes. % which could aid in the survivors' goal of achieving dignified reintegration.

\begin{table*}[]
\caption{Our PD configurations emerged from the recognition of the survivors' dependency on the organizations for support as well as the constraints on the organization.}
\label{tab:configurations}
\begin{tabular}{@{}l|l@{}}
\toprule
\textbf{Structural Constraints and Dependency} & \textbf{Action for Empowerment} \\ \midrule
Mutual dependency & Presenting crafting and mutual bond as assets \\
Technological aspirations & Operationalizing assets using technology \\
Dependence on donor funds & Cultivating mutual support mechanisms \\ \bottomrule
\end{tabular}
\end{table*}

\subsection{Configuration 1: Understanding and Working With Mutual Dependency}

This configuration re-conceptualizes the relationship between survivors and organization as one of \emph{mutual} dependence. 
This re-conceptualization led us to slowly question deeply held beliefs about the survivors' deficits, and to notice the possibilities inherent in the survivors' existing assets: their knowledge of creating local handicrafts and their close bond with one another. 

\subsubsection{Mutual Dependence}

In our initial ethnographic study, we noted the sister-survivors' dependence on SO. 
In addition to a safe living environment, SO provided psycho-social counseling and training programs to help with future employment.
These services are some of the few resources available to survivors in their reintegration journey.
Both PO and SO appeared to be doing their best with the resources they had, and many staff members, particularly at SO, reported that the opportunity to work on problems that had affected them was a significant motivator for their work. 
% Yet their beliefs and practices positioned the survivors as passive recipients of support. The survivors' deficits and needs were highlighted in many forms and mediums such as in their annual report and grant applications. 
%We came to reconceptualize this relationship as one of mutual dependency. 
At the same time, the organizations were reliant on the survivors in the sense that they needed the survivors to rely on their services to justify their work and demonstrate their impact to donor organizations and others.
The needs of the survivors were highlighted in a variety of forms and mediums, such as the annual reports and grant applications.
For instance, SO required survivors to recollect their past so that their stories could be used in official documents and grant proposals. 
The sister-survivors found this painful. One of them shared with us, ``\textit{I had already forgotten it} [the past events] \textit{and being reminded of it was hard. I was so sad for 2-3 days. We have left that place and moved on.}''
But, as one staff member remarked that they have ``stories to rattle the hearts of everyone in the US''.

The actual difficulties of the sister-survivors, the staff's own good intentions plus externally-driven incentives to position the sister-survivors' as needy operated together to make it difficult for the staff to focus on the ways in which the sister survivors were strong.

\subsubsection{Uncovering Assets That Align with the Mutual Dependence}

Taking sides with the survivors necessitated challenging the prevalent deficit perspective. 
However, mutual dependence present in the setting required us to carefully and incrementally configure our approach, first uncovering the assets available to the sister-survivors and then positioning them as resources to both the survivors (e.g., in achieving dignified reintegration) and the organization (e.g., in future grant proposals).

\begin{figure*}
\includegraphics[height=2.8in]{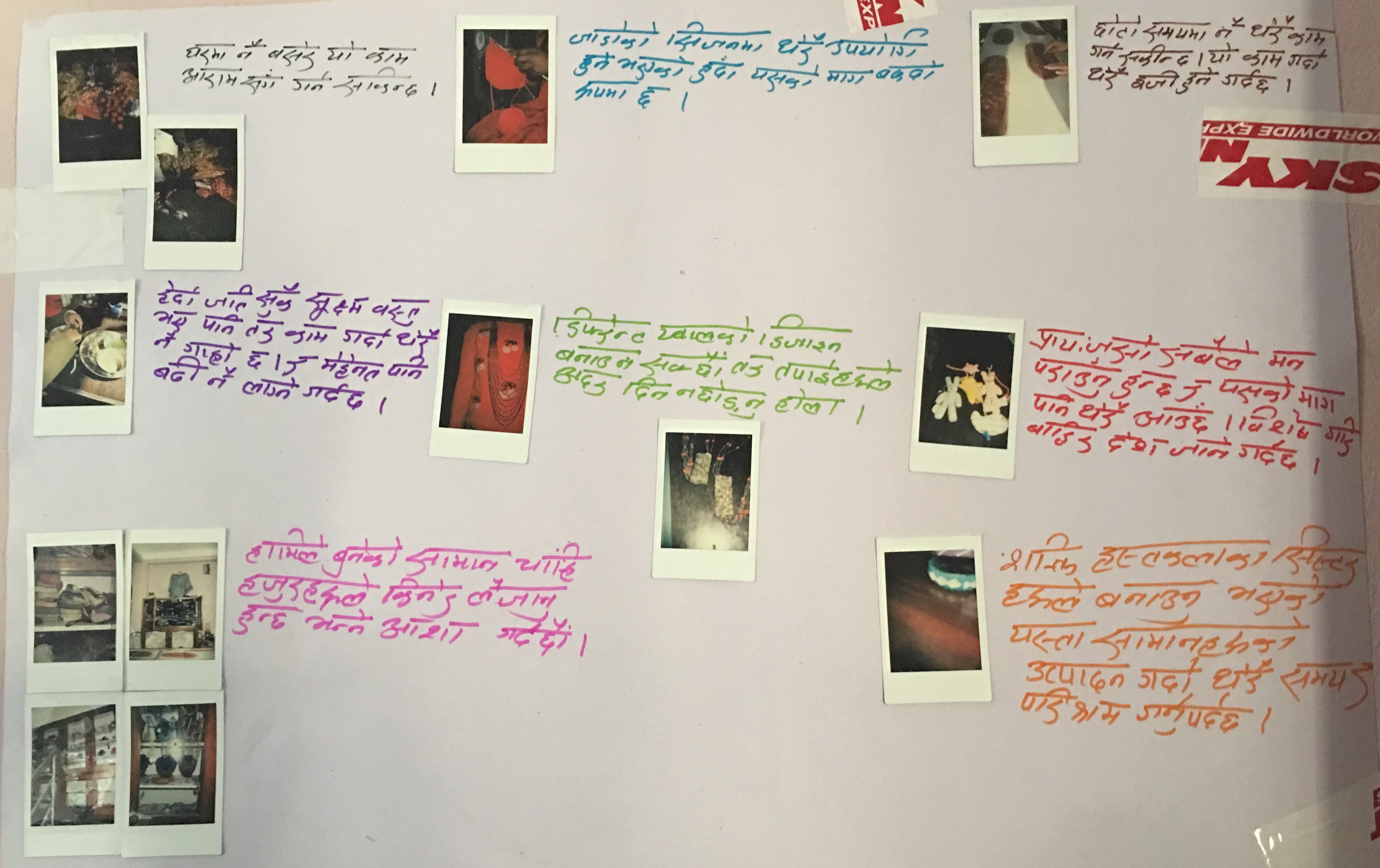}
\centering
\caption{The two posters created during the social photo-elicitation session were placed on SO's office walls, showcasing sister-survivors' work for the staff members and visitors.}
\label{fig:posters_on_SO_walls}
\end{figure*}

To this end, we designed a photo-elicitation activity in which the sister-survivors took photographs of their surroundings and then discussed them two days later.
We held two rounds of group discussions based on the photographs.
The sister-survivors shared what the photos meant to them and collectively came up with a summary text to accompany the photo on a poster (see Figure \ref{fig:posters_on_SO_walls}). 
We posted the posters on SO’s walls, creating an opportunity for the sister-survivors to showcase their achievement to the staff members. 
The posters are still on SO's walls.

The discussions shed light on various aspects of the sister-survivors' lives in the shelter home as well as their future visions, including the economic and social challenges they anticipate after leaving the shelter home.

We identified two primary assets available to the survivors living in the shelter home based on our discussions: their knowledge of creating local handicrafts and their mutual bonds with one another. 
Both of these assets were fragile. 
The sister-survivors were aware that sales of handicrafts in the local market was declining. Some found crafting boring and difficult. 
While the sister-survivors valued their mutual bond, they were concerned about the possibility of losing touch once they left the shelter home and moved across the country.

\paragraph{Knowledge of crafting as an asset:}
We balance the goal of empowering the survivors while attending to the mutual dependence by emphasizing the sister-survivors' knowledge of crafting as an asset.
The sister-survivors saw crafting as a valuable skill they had worked hard to learn in the shelter home.
Most of them saw it as a potential source of livelihood in the future. 
It was also a source of income in the present, as they received 25\% of the sales proceeds.
They wanted to showcase their crafts, thought that crafting was therapeutic, and saw the the potential for using crafts to bring their family together. 
We hoped to position survivors as agentic actors with the power to achieve dignified reintegration by emphasizing crafting as a valuable asset that was already available to them.
From SO's perspective, they had invested in setting up the handicraft workshop, hiring specialized trainers, and renting a storefront.
SO generated some revenue through handicraft sales and, from an operational standpoint, provided survivors with attainable and accessible skills-based training.

\paragraph{Mutual bond as an asset:}
Sister-survivors had very few opportunities to build connections with people outside of the shelter home. Some of them were shunned by their birth families.  However, the shared living situation, their awareness of each others’ past, the shared work, and the mutual support they provided each other fostered a strong bond between them.  The mutual bond was talked about as family-like reliance. Their mutual bond could be beneficial for long-term emotional support and collective action. From SO's perspective, the survivors’ mutual bond could be beneficial in facilitating their reintegration away from their hometowns. Moreover, promoting mutual bonds does not require additional financial investment. 

Despite problems, the sister-survivors saw the value of the two assets in their daily lives and could envision futures with them. 
Critically, the assets were enabled by the processes put in place by SO, so our next steps could be configured to align with SO's priorities.

\subsubsection{Presenting the Identified Assets to the Staff Members}
Power is exercised through decision-making \cite{bratteteig2012disentangling}.
A space for making moves to empower survivors entailed paying attention to how we included staff members and sister-survivors in decision-making.
In particular, the sister-survivors' dependence on the organization meant that we had to approach the sister-survivors and the staff in a layered way. 

% We involved staff members in decision-making throughout our engagement, sharing our findings and charting next steps.
% Staff members were also involved in the iterative design of approaches.
% Once designed, we proposed the idea to a group of sister-survivors whose involvement facilitated further iterations.  
% This enabled us to handle the power disparity while gaining organizational buy-in and placing the sister survivors as agentic actors. 

For example, after identifying the initial assets, we shared our findings with the organization.
Following that, three staff members took part in a conversation about devising techniques to build upon the assets.
The staff members suggested providing photography and photo editing lessons so that survivors could market their handicrafts.
Other suggestions included developing a system for the survivors to report incidents to the police and developing a mobile application to learn new skills.
A technology component was evident in all of these suggestions. 
To mitigate the chances of the staff members envisioning technology as a charismatic panacea \cite{ames2019charisma, toyama2015geek}, we emphasized the importance of using technology as a means rather than an end, specifically to build upon the sister-survivors existing assets towards broader possibilities. 
\subsection{Configuration 2: Operationalizing Assets to Highlight Alternatives}

The sister-survivors' knowledge of crafting and their close bond with one another were assets that could be leveraged in the context of mutual dependence.
As we sought to build on these assets towards future possibilities, we saw opportunity in the organization's technological aspirations which they saw as a critical component in our continued co-engagement.  
This led us to explore the possibility of building on the assets through a tailored web application and associated workshop that we incrementally extended to collectively explore broader possibilities and alternatives.

\subsubsection{Emphasizing Assets While Leveraging Technology}

The value SO put on technology meshed with the sister-survivors' apparent interest in learning. However, there was a risk of perpetuating dependency and eroding their agency if we positioned technology---which is an external resource---as the central element in empowering the sister-survivors. 
Indeed, the organization's prior experience with technology introduction had done just this. 
Twelve sister-survivors had been taught Photoshop during a two-week introduction. They found the experience overwhelming and reported that they did not learn much. One sister-survivor shared, ``...\textit{ when we were learning to edit} [photos], \textit{we had never touched a laptop before to know anything.}''

Considering this, our configuration emphasized the importance of using technology as a means rather than an end, specifically by defining what assets are, labeling their assets, and then presenting technology as a means to further those assets.

\paragraph{Using technology to build on assets:}

We introduced a voice-annotated web application called Hamrokala (``Our Craft'' in Nepali) through a ten-day workshop. 
The web application supported the sister-survivors to post crafting items as if for sale on the internet.
The setting was made communal by the sister-survivors’ behavior but the web application was also tailored to build on their communal orientation, allowing activities such as drawing and sharing design ideas, seeing other sister-survivors’ crafts, and commenting on them. 
Further, the voice annotation served to relieve the pressure to read while also making a positive contribution to the sociality of the activity by making individual computer actions more public. 

Initially, we conducted the workshop with a group of nine sister-survivors.
%At the end of the workshop, the sister-survivors were paired with a staff member to whom they showcased the digital artifacts they had created and the digital skills they had learned. 
We later replicated and extended the workshop with a group of ten sister-survivors. 
We extended it in three ways. 
First, we expanded on the technical skills and the higher literacy level of that cohort to incrementally introduce widely available systems like Google Search and Wikipedia. 
Hamrokala was a tailored application that would not be present outside the shelter home so we explored ways to build upon the computing skills developed through Hamrokala towards more general systems. 
Second, we facilitated the sister-survivors in using technology to find out more about the world. 
Third, we designed activities to support the sister-survivors to chart avenues for interaction with societal actors and institutions.

The sister-survivors were able to use computers and seemed to enjoy them. 
They worked together to learn, explore, and troubleshoot technology. They formed new rules for the space and in the end, expressed a desire to learn more about computing.
When we extended the session, the sister-survivors appropriated technology to learn English and valued the opportunity to help others such as by contributing to Nepali Wikipedia pages. 
Within a playful space, they examined computing's potential in gaining information about public services (e.g., documents required to obtain a citizenship certificate or open a bank account) as well as its limitations (e.g., why only photos of a national park showed up when they searched for their hometown).

Importantly, the technological exploration enabled a space to examine the potential and limitations of the assets available to them.
During the Hamrokala workshops, the sister-survivors discussed such matters as the effort required to make a handicraft, the selling prices, and different techniques to market the crafts. 
The activities fostered discussion of alternative pathways. 
Some of the sister-survivors anticipated that further learning with computers could lead to ``office jobs''. 
Using Wikipedia, they discovered other career paths that they could take. 
This led them to see the potential of pursuing higher education to become accountants, lawyers, social workers, or veterinarians; none of those visions involved crafting.

\begin{figure}
    \centering
    \begin{subfigure}{0.35\linewidth}
        \centering
        \includegraphics[height= 6.5cm]{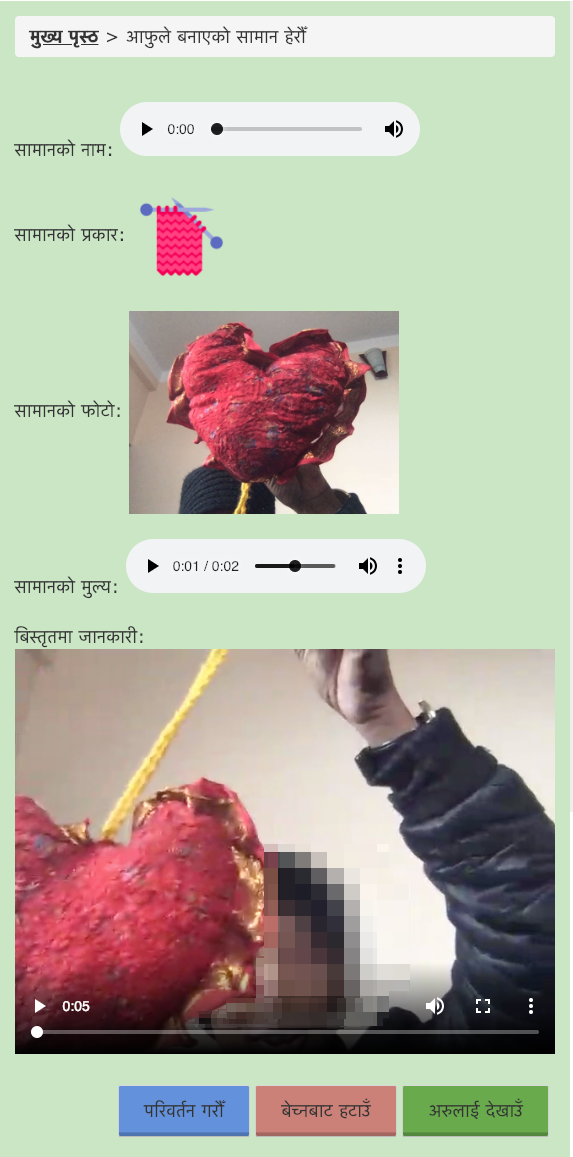}
        \caption{Example craft posted on Hamrokala}
    \end{subfigure}
    \hfill
    \begin{subfigure}{0.45\linewidth}
        \centering
        \includegraphics[width=\linewidth]{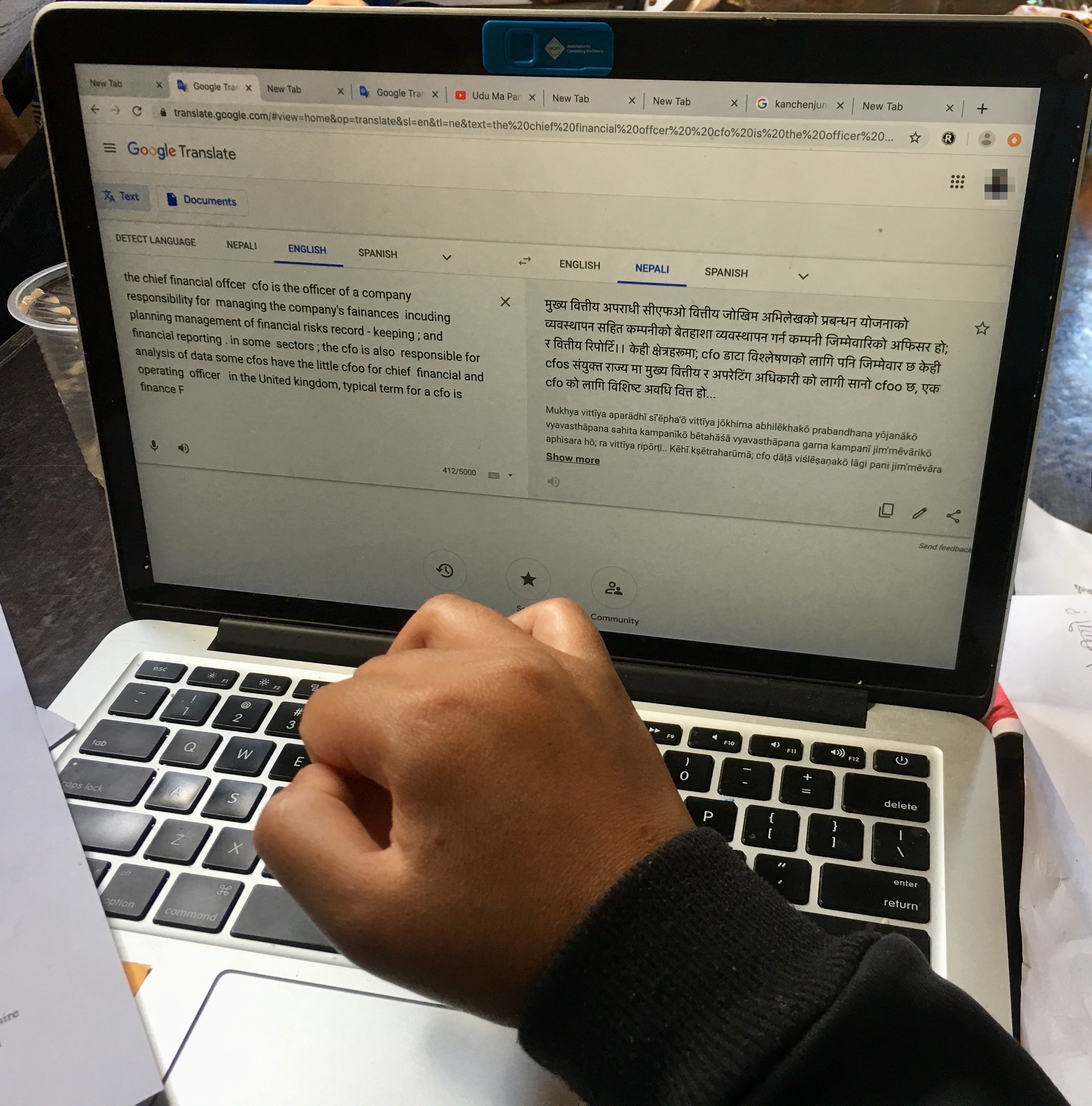}
        \caption{A sister-survivor translating the Wikipedia page on chief financial officer}
    \end{subfigure}
    \caption{We began by uncovering and reflection on the assets available to the sister-survivors. (a) We then built on the identified assets using a tailored web application. (b) We extended the exploration to introduce widely available systems such as Google Search and Google Translate, and Wikipedia.}
    \label{fig:addNewItem}
\end{figure}

\subsubsection{Showcasing Assets to Challenge Deficit Perspective}

On the last day of the Hamrokala workshop, we designed an activity where the sister-survivors showed their computing skills to the staff members. We were aware from our previous interactions that the staff members valued computing skills. At the same time, SO’s prior failure introducing computers had eroded the staff members’ confidence that the sister-survivors would be able to learn computers. By demonstrating that they successfully learned and adapted computers, the sister-survivors were able to challenge the deficiency-focused belief around them, exemplified in the warden's exclamation, ``\textit{they learned all this in two weeks!}''
In fact, the sister-survivors’ successful use of technology prompted the staff members to propose the possibility of training survivors to become computer trainers. It indicates that the approach mitigated some of the discourse around the survivors’ deficits.
\subsection{Configuration 3: Leveraging Assets to Promote Possibilities of Mutual Support}

In our discussions with staff members, we highlighted how the sister-survivors' assets and feedback informed the design of our approaches including the design of technology. 
We also added steps for the sister-survivors to showcase their assets to themselves, one another, and the staff members. 
%For example, the posters created by the sister-survivors during the photo-elicitation sessions were displayed on the organization's walls and later the sister-survivors demonstrated their technical skills to the staff members. 
These steps helped challenge the prevalent belief in the survivors' deficiencies. 
Indeed, in the staff members' proposition to train survivors in becoming computer trainers, we heard an appreciative assets-first perspective that saw survivors as agentic actors with valuable skills. 
Buoyed by the positive appreciation of the survivors' assets, we began inquiring about changing the workplace practices by including the sister-survivors in the organization's project design process. 
Project design involves designing intervention programs for survivors and proposing the project to donor organizations. 
Only projects selected for funding eventually get implemented. 
To date, none of the survivors---including survivors who are part of the organization's staff---have been involved in designing projects. 

As we began inquiring about the possibility of centering the sister-survivors' voice by involving them in the organization's project design process, we encountered a barrier emerging from the organization's dependency on donor funds. 
We thus sought to configure our approach, carefully attending to the dependency while also pushing for the survivors to have a greater voice in the project design process.

\subsubsection{Dependence on Donor Funds}

In 2016, 60.7\% of SO's annual funds came from donor organizations, most of which were international agencies. 
Given the dependency, SO's programs are influenced by the priorities established by the donor organizations whose agenda is often influenced by the US Department of State’s annual Trafficking in Persons (TIP) report, which provides an account of trafficking issues prevalent in the majority of the countries across the globe. As one staff member reported, ``\textit{TIP report always has recommendations at the end of the section for each country. There are recommendations for Asia too, for South Asia, and within it for Nepal. We have to shuffle through detailed and nuanced information. After gathering the information, we write the proposal. Those that are written like that have a higher chance of getting in} [funding].''
 
The difference in tasks---between writing grant proposals and implementing the funded projects---seems to have enforced a hierarchy between staff members who are professionals (who were referred to as ``technical-staff'') and survivors who were part of the staff team (referred to as ``member-staff''). 
Technical-staff manages the project design and proposal work, with limited involvement of the member-staff or the survivors:
\begin{quote}
\textit{We} [technical staff] \textit{are the ones who write the programs at SO...In the first design phase, in all the various drafts that are created, we are there. Until the stage when the grant is approved, we are the ones doing it.}
\end{quote}

As we inquired into understanding the barriers to including the member staff and the survivors in the project design process, we heard a range of institutional and organizational factors.
Individual deficiencies such as lack of education or critical thinking were mentioned as significant barriers, as heard in a staff member's comment, ``\textit{All people at all levels to come to a single table to discuss is not possible. People understand according to their levels and they talk according to their level. In writing projects, they have to have some critical thinking abilities, someone who has understood a little bit.}''
Despite appreciating the sister-survivors' technical know-how, the staff member's deficit-centered belief about the survivors was not completely erased. 
One staff member exclaimed, ``\textit{If you ask the sisters} [survivors] \textit{they will say `we don’t know anything'.}''

The staff members also hinted at SO's precarious financial position which made it risky to try new approaches. 
For example, one staff member shared, ``\textit{We lack in core funding so we can't say no to projects. We are needy a lot.}'' 
SO's reliance on external donors is a reality of the ground; changing the structure of SO's dependence on donor funds is challenging and beyond the current scope of our influence.

\begin{figure*}
\centering
\includegraphics[height=2.8in]{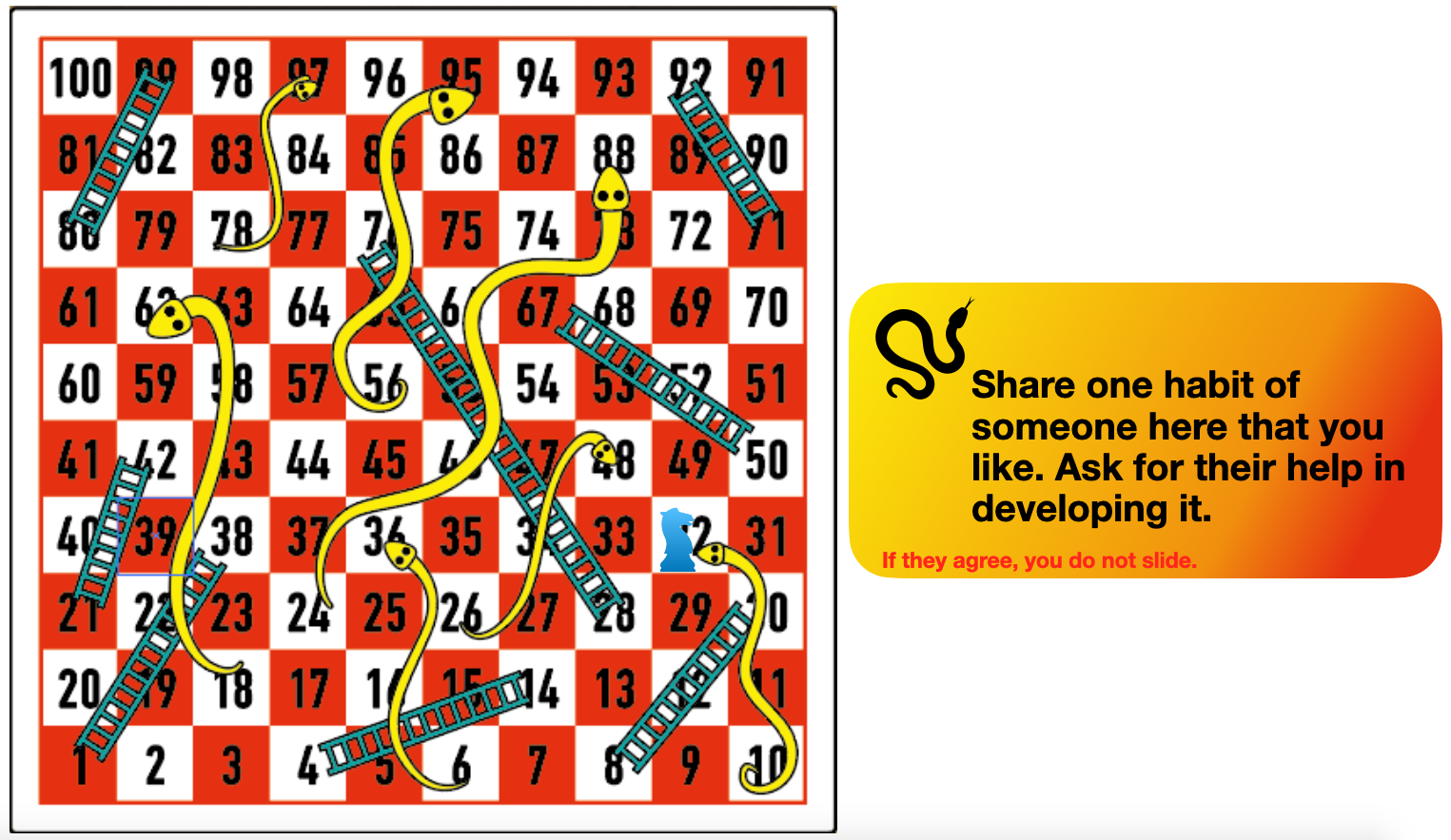}
\caption{Players take turn to roll a dice with the goal of reaching the end (cell number 100. If a player lands on a cell with a snake or a ladder, they choose a card from a corresponding deck.
The ladder card includes a personal question about their lives in the shelter home or their future vision.
The snake cards pose relational questions involving other players and can collectively decide to save the player from sliding to a lower cell. Before the game, the group can change the questions.}
\label{fig:snakes_and_ladder}
\end{figure*}

\subsubsection{Playful Interaction to Showcase Knowledge}

Acknowledging SO's reliance on donor funds and hesitance to change their project design process but at the same time noting the importance of taking sides with the sister-survivors so that they have a voice in the design of the projects in the future, we have begun co-designing a playful activity that aims to facilitate opportunities for the sister-survivors to express themselves and share critical thoughts or values that may inform project design. 
The game involves an adaptation of Snakes and Ladders, a popular game in Nepal. 
The modification involves adding collaborative elements where players can connect, share knowledge, and appreciate other players' strengths (see Figure \ref{fig:snakes_and_ladder}). 
We are currently working with two staff members in the organization to iterate on the initial set of questions.

We hypothesize that the questions and the discussion can support the sister-survivors in slowly sharing and contributing to project design without significantly disrupting the current process. 
The playful nature of the game reduces the pressure on the sister-survivors to deliver something profound. Likewise, it facilitates an opportunity for the staff members to learn from and about the sister-survivors without adding a significant amount of work. 

We plan to present the game to the staff members and the sister-survivors to play together.
While the game has not yet been implemented, the staff members have accepted it and seem to see value in using it in the organization. 
The acceptance indicates a small change in the space where the sister-survivors' knowledge and other assets are appreciated.
We contend that while it is a small change, it is a significant one considering the various constraints on the organization and the sister-survivors.
\section{Discussion}

Three configurations emerged from acknowledging that the services the survivors receive from the anti-trafficking organization, while limited and problematic, are valuable to their reintegration journey. 
The first configuration was shaped by the perception of mutuality of dependence between  the sister-survivors and the organization.
The mechanism for empowering participation then became to focus upon and highlight pre-existing assets of the sister-survivors rather than engaging in a wholesale redesign.  
The second configuration arose because technological knowledge and adoption were so highly valued within the organization that, if designed with enough care to be usable, they could be utilized to change how sister-survivors were perceived both by themselves and by the organization.  
The third configuration is based on the first two; demonstrating that the sister-survivors have agency with technology seemed to promise to the organization that their voices could also have something to contribute to the conceptualization of future interventions. 

We do not yet know whether our proposed intervention is ``in the zone'' where it can work, but the idea is to empower the sister-survivors in a way that makes the organizations' dependency on them both more visible and more dignified. 
To be ``in the zone'' where it can work, the voices must be sufficiently loud and clear and yet not threatening. 
Success in the design of each of these interventions means finding a way to take action that fulfills all of the constraints of the situation well enough and still promotes empowerment.  
It means taking sides, but not too much. 
Success should therefore be judged not in absolute terms, but in how it resolves the design tensions inherent in the context \cite{tatar2007design}.   

The constraints and the dependency essentially involve enactments of power that, in some cases, transcend the locus of the survivors and the organization, and may involve larger entities like the government and international donor organizations. 
In undertaking a transformative agenda, we must recognize our limited scope as we have to engage with the layers of power differentials, identities and historicity that comprise the realities of the ground \cite{massey1995thinking}.
Assets-orientation enabled us to make limited but valuable moves within the constraints. 
While the scope of our action is---and will be---limited, by attending to the position of the vulnerable population within existing power structures, we can frame incremental moves towards transformation.

In particular, emphasis on sister-survivors’ assets positions them not as powerless individuals but as agentic actors who can contribute to the organization and the broader society. 
Once this pathway is established, i.e., through Hamrokala, it becomes easier to broaden the prevalent view to include less tangible assets. 
We intend to gradually bring change in the power dynamics, as exemplified by our collective exploration of using Snakes and Ladders to include sister-survivors in the project design process. 
Each of our moves constitutes a small step in a long-term transformation.

\subsection{Bridging Agonistic Approach}

Politics matter \cite{balka2010broadening, beck2002p}.  
They matter in a context like ours where, to realize democratic control, PD practitioners need to take sides with the vulnerable population. 
Being political requires that PD practitioners form relationships and use their power in such a way that the voices of the vulnerable group are prioritized \cite{karasti2010taking, balka2010broadening, akama2018practices, kendall2018disentangling}. 
In the first configuration, we used our position to insist on acknowledging the mutuality of dependence. 
In the second configuration, we leveraged our technical skills and position to focus on assets. 
Explicitly focusing on assets was a political move, one that empowered the participants within the existing structure.
In our third configuration, we used our position to show value in including the survivors in a core organizational process, that of designing projects. 
Taking sides inherently involves different degrees of confrontation \cite{bjorgvinsson2012agonistic, bodker2018participatory}.

Scholars posit that change necessitates an agonistic space in which heterogeneous perspectives can be expressed and contested, enabling a realization that ``things could always have been otherwise and every order is predicated on the exclusion of other possibilities'' \cite[pp. 549]{mouffe2009democracy} (see also \cite{bjorgvinsson2012agonistic, hillgren2016counter, disalvo2015adversarial}).
We endorse this belief.
However, as noted above (Section \ref{sec:bridge}) enabling a space for contestation requires consideration of local context and power dynamics. %, ensuring that there is not a significant power difference between actors involved and that the actors do not face repercussions. 
We attend to the sensitive nature of this setting by \emph{not} inviting the ``profound conflicts'' that Bødker and Kyng argue for \cite{bodker2018participatory}.
Reflecting on our three configurations, we identify three mechanisms that allowed us to balance the design tensions inherent in the setting.

\subsubsection{Highlighting Interdependency to Promote Collaborative Entanglement} %\subsection{Highlighting Interdependency Between the Actors}

One tactic of agonism involves revealing the hegemony in the setting \cite{disalvo2015adversarial}. 
In contexts of dependency, revealing the hegemony may not be acceptable or lead to actionable change. 
In our case, from one perspective, the organization's beliefs and practices could be seen as hegemonic. 
Indeed, it is dominant; the programs implemented by the organization shape the survivors' possibilities.  
Yet, from another perspective, the organization's practices arise from a place of care that can be sustaining as well as conflicting and restrictive \cite{murphy2015unsettling, de2017matters}.
For instance, the organizations' internal practices are deeply rooted both in beliefs about the survivors' deficiencies and in values of protection and care. 
Revealing the dominance of the organization would not attend to these conflicting values. 
Moreover, in structures of dependency, the adversaries are not equal in terms of power and agency; the dependent group may lack the agency to contest, and the more powerful group may not value the contestation. 

Focus on the dependent groups' assets can illuminate the interdependency between the groups, allowing us to establish a ``collaborative entanglement'' that shifts actors from a direct adversarial position toward interdependent exploration of potential futures.
Critically, collaborative entanglement complicates the us-and-them perspective that is central to adversarial stances \cite{mouffe2005some, disalvo2015adversarial}. %reduce the emphasis on?

Elements of collaborative entanglement in our work include acknowledging the assets' values to our research agenda and the organization's goals; engaging the staff members in planning how to build on the assets; and  
using the assets to create a space in which the sister-survivors were seen as individuals with strengths to contribute to the organization.
These incremental steps towards emphasizing interdependence were critical in gaining traction and acceptance for change.

\subsubsection{Attending to Contingent Factors to Showcase Broader Possibilities}

Building on Suchman \cite{suchman2006human}, DiSalvo proposes an agonistic tactic of reconfiguring the remainder in which we ``include what is commonly excluded, giving it privilege, and making it the dominant character of the designed thing''  \cite[pp. 64]{disalvo2015adversarial}. Focusing on the sister-survivors' assets does this, because it challenges the deficit approach;
however, a more nuanced perspective is required. 
We contend that what has been excluded or ignored is highly contingent, and that configuring its inclusion requires careful consideration to avoid reinforcing existing power differences.
In particular, the reconfiguration must enable the actors to see broader alternative possibilities.

For instance, initially, we identified the sister-survivors' crafting skills as an asset. 
Crafting is highly contingent on the infrastructure made available by the organization and reconfiguring it as a central part of the sister-survivors' reintegration journey could have increased their dependency on the organization while also limiting the possibilities they saw for themselves. 
Instead, we built upon their crafting skills to introduce technology, whose knowledge in and of itself became an asset to the sister-survivors. 
The introduction of technology supported the sister-survivors in considering alternative futures beyond those presented to them through the organization's services (e.g., moving away from seeing only crafting as a source of livelihood).  
Indeed, the sister-survivors' exploration of alternative careers, as well as the organization's plans to train the sister-survivors to become computer trainers, indicate that our careful inclusion of the assets was successful in creating space to envision alternative possibilities.

\subsubsection{Emphasizing a Provisional Collective}

Supporting a collective capable of ``disarticulating the existing order'' \cite[pp. 109]{disalvo2015adversarial} is an agonistic design strategy. PD practitioners can play an important role in forming a collective that can act on matters of concern to them \cite{dantec2013infrastructuring}. 
In fact, a commitment to assets-based design entails forming a collective that appreciates and engages with the assets available to them \cite{wong-villacres2021Reflections}. 
However, how the collective is formed and positioned requires careful consideration of   
the structure of dependency.
A narrow formulation of collectivity %---that is of a narrower notion of who the ``us'' is in an adversarial stance \cite{mouffe2005some, disalvo2015adversarial}---
may result in the exclusion of actors who are, on some level, aligned with the overall objective of the intervention. %Their inclusion may result in the collective gaining more power.
A rigid conceptualization may fail to account for uncertainties.
PD practitioners must engage in an ongoing process of forming and re-forming the collective. 

While we saw the need for internal changes, we also saw the organization as a valuable resource for the sister-survivors, possibly beyond their lives in the shelter home. 
Considering this, we included the staff members in the collective by involving them in decision making throughout our engagement.
We challenged the staff-members' deficit-focused perspectives by emphasizing the sister-survivors' assets but, importantly, without excluding them. 
This allowed us to gain organizational buy-in to the extent where we could propose a different collective, one in which the survivors could be involved in the organization's core functions. 

At the heart of agonism lies the belief that democratic space requires diverse, heterogeneous, pluralistic perspectives to be expressed and contested. 
We believe that this is critical in realizing change.
However, in some contexts, such as within structures of dependency, agonistic spaces with an adversarial stance may not be appropriate. 
In such contexts, PD practitioners can leverage their position and shape PD engagement to draw out pluralistic perspectives and broader possibilities, thereby forming the necessary infrastructure that can eventually manifest in an agonistic space.

\section{Conclusion}

Our five-year-long engagement with an anti-trafficking organization in Nepal and survivors of sex trafficking supported by the organization aimed to support the survivors in achieving, what they call, dignified reintegration back into Nepali society. 
We built a relationship with the organization based on trust and mutuality, and leveraged our position to configure approaches that balanced multiple constraints and dependency present in the setting.
In particular, our approach aims to illuminate a sustainable pathway to appreciate the sister-survivor's assets, and, thereby, bring about change in individual beliefs and organizational processes.
%Kendall and Dearden ask ``Who participates, with whom, in what, and why?'' \cite[pp. 7]{kendall2018disentangling}.

Through this paper, we aim to deepen PD's exploration of embracing politics and agonistic approaches in sensitive settings. 
Considering structures in which various vulnerable population find themselves, we contend that while agonism is important, we need to complicate the picture by thinking about collaborative entanglements, contingencies, and provisional collectives.

\bibliographystyle{ACM-Reference-Format}
\bibliography{sample-base}

\end{document}